\documentclass[a4paper,11pt]{article}
\newcommand{\tr}{\,\mbox{Tr}\,}
\newcommand{\lesssim}{\:\mbox{\raisebox{-3pt}{$\stackrel%
{\displaystyle <}{\sim}$}}\:}
\begin{document}
\begin{titlepage}
\begin{flushright}
UWThPh-1999-36\\
May 1999
\end{flushright}
\vspace{1cm}
\begin{center}
\Large
\textbf{A 4-neutrino model with a Higgs triplet}\\
\vspace{1cm}
\normalsize
\textit{W. Grimus, R. Pfeiffer, T. Schwetz}\\
\textit{Institute for Theoretical Physics, University of Vienna,}\\ 
\textit{Boltzmanngasse 5, A--1090 Vienna, Austria}\\
\vspace{1cm}
\textbf{Abstract}\\[3mm]
\begin{minipage}{0.8\textwidth}
We take as a starting point the Gelmini -- Roncadelli model enlarged by
a term with explicit lepton number violation in the Higgs potential and add a
neutrino singlet field coupled via a scalar doublet
to the usual leptons. This scenario allows us to take into
account all three present indications in favour of neutrino
oscillations provided by the solar, atmospheric and LSND neutrino
oscillation experiments. Furthermore, it suggests a model which reproduces
naturally one of the two 4-neutrino mass spectra favoured by the
data. In this model the solar neutrino problem is solved by large
mixing MSW $\nu_e\to\nu_\tau$ transitions and the atmospheric neutrino
problem by transitions of $\nu_\mu$ into a sterile neutrino.\\[3mm]
\end{minipage}
\end{center}
\end{titlepage}

\section{Introduction}
\label{intro}
At present there are three indications in favour of neutrino
oscillations with three different scales of the corresponding 
neutrino mass-squared differences. Taking into account that in the LEP
experiment the number of light active neutrinos was determined to be
three, it follows that at least one sterile neutrino is required 
to describe all present neutrino oscillation data (for reviews see, e.g.,
Refs.~\cite{valle,conrad,BGG98,mohapatra}). In the following we
confine ourselves to the 4-neutrino case which was discussed in
many papers for a number of reasons (for an incomplete list see
Ref.~\cite{four}). From present
experimental data the nature of the 4-neutrino mass spectrum can be
inferred \cite{BGG96,OY97,barger}
and also information on the $4 \times 4$ unitary
neutrino mixing matrix $U$, which is defined by
\begin{equation}\label{mixing}
\nu_{\alpha L} = \sum_{j=1}^4 U_{\alpha j} \nu_{jL} 
\quad \mbox{with} \quad \alpha = e, \mu, \tau, s 
\end{equation}
can be obtained.
In this relation, $\nu_{\alpha L}$ denotes the fields with definite
flavours or types whereas $\nu_{jL}$ denotes the left-handed part of
the neutrino mass eigenfields. The measurement of the 
up-down asymmetry of the atmospheric muon neutrino flux \cite{super-kam} 
allows to draw definite conclusions on the types of possible neutrino
mass spectra \cite{BGG99}
for the whole range of the mass-squared difference $\Delta m^2_\mathrm{LSND}$
determined by the LSND experiment \cite{LSND} and other short-baseline
neutrino oscillation experiments. In this way only two types
of mass spectra with two pairs of close masses are allowed. These mass
spectra can be
characterized in the following way \cite{BGG96,BGG99}:
\begin{eqnarray}
\mbox{(A)} &&
\quad
\underbrace{
\overbrace{m_1 < m_2}^{\mathrm{atm}}
\ll
\overbrace{m_3 < m_4}^{\mathrm{solar}}
}_{\mathrm{LSND}} \;, \nonumber \\
\mbox{(B)} &&
\quad
\underbrace{
\overbrace{m_1 < m_2}^{\mathrm{solar}}
\ll
\overbrace{m_3 < m_4}^{\mathrm{atm}}
}_{\mathrm{LSND}}
\;.
\label{AB}
\end{eqnarray}

The task of accommodating a light sterile neutrino in an extension of
the Standard Model poses serious problems to model builders. In
particular, it seems difficult to reconcile the mass spectra
(\ref{AB}) and the large mixing observed in atmospheric neutrino
oscillations with the original see-saw mechanism
\cite{seesaw}. However, models have been proposed exploiting the
``singular see-saw mechanism'' \cite{singular}
which naturally achieve a large active -- sterile neutrino mixing
\cite{CKL,chikira,liu}. Since a large mixing angle $\nu_e\to\nu_s$
transition as a solution of the solar neutrino puzzle is not
compatible with the solar neutrino data \cite{petcov}, the
``singular see-saw mechanism'' offers the possibility to explain the
atmospheric neutrino anomaly by $\nu_\mu\to\nu_s$ oscillations.

A large active -- sterile neutrino mixing seems to be excluded by
big-bang nucleosynthesis if only less than 4 effective light neutrino
degrees of freedom ($N_\nu$) are allowed
(see Refs.~\cite{shi,OY97,BGGS98} and citations therein). 
However, the upper bound on $N_\nu$ depends, in particular, on the
primordial deuterium abundance $(D/H)_P$ for which conflicting
measurements exist. For the low value of $(D/H)_P$ the value of
$N_\nu$ should rather be close to 3 \cite{burles} 
whereas a high ratio $(D/H)_P$ allows also values of
$N_\nu$ around 4 \cite{olive}. In the following we adopt the
hypothesis that $N_\nu = 4$ is allowed.

In this paper our starting point to construct a 4-neu\-trino model is
not the singular see-saw mechanism but an extension of the Standard
Model in the scalar sector. Nevertheless, we will see that one can
arrive at a scenario equivalent to the one obtained in
Ref.~\cite{CKL}. The possible scalar multiplets extending
the Standard Model are simply obtained
by studying the representations of SU(2)$\times$U(1) contained in all
the fermionic bilinears which can be formed. Apart from the
scalar doublet there are only three possibilities: a triplet, a singlet with
charge +1 and a singlet with charge +2 \cite{kummer}. 
The basic and most prominent models founded upon these scalar 
multiplets are given by the models of
Gelmini -- Roncadelli (GR) \cite{GR}, Zee \cite{zee}  and Babu \cite{babu},
respectively, with Majorana neutrino masses at the tree, 1-loop and 2-loop
level. Our discussion is based on the GR model. In its original version
\cite{GR} it possesses a spontaneously broken lepton number leading to a 
majoron and a light neutral scalar such that the $Z^0$ vector boson decay
into these two scalars has a width of twice the decay width of 
$Z^0 \to \nu_\alpha \bar \nu_\alpha$ where $\nu_\alpha$ denotes any of the
three active neutrinos \cite{georgi}. Since there is no room for such a decay
according to the LEP measurements, we explicitly break the lepton number by a
cubic term in the Higgs potential (see, e.g., Ref.~\cite{ma}) in order
to make the majoron heavy. The vacuum
expectation value (VEV) of the neutral member of the Higgs triplet gives a
Majorana mass matrix at the tree level for the active neutrinos. To
incorporate a sterile neutrino singlet field $\nu_{sR}$ we couple it to the
Standard Model lepton doublets via a Higgs doublet (for an analogous procedure
in the framework of the Zee model see Ref.~\cite{tanimoto}) 
and invoke a symmetry to
forbid the mass term $\nu_{sR}^T C^{-1} \nu_{sR}$ where $C$ is the charge
conjugation matrix. The main point of our scenario is to exploit the relation
\begin{equation}\label{vev-rel}
|v_T| \ll v \,,
\end{equation}
where $v_T$ is the VEV of the Higgs triplet and $v$ denotes the largest
absolute value of the VEVs of the scalar doublets. A large triplet VEV would
destroy the tree-level relation $M_W = M_Z \cos \theta_W$ between the $W$ and
$Z^0$ boson masses and the Wein\-berg angle and 
the precision measurements place
a stringent bound on $v_T$ \cite{erler}. With the two scales $v$ and
$v_T$ we will show that at this stage we have a model equivalent to the one
described in Ref.~\cite{CKL}. Finally, we will introduce a discrete 
symmetry to achieve maximal $\nu_\mu$--$\nu_s$ mixing, to some extent without
fine-tuning. In the final stage of our model we will have three scalar
doublets in addition to the Higgs triplet.

Other 4-neutrino models with Higgs triplets have been considered in
Ref.~\cite{triplet}. 

The paper is organized as follows. In Section \ref{Gelmini-Roncadelli} we will
present a thorough discussion of the GR model with explicit lepton
number violation since this model is the basis of the further
discussion in the paper. The sterile
neutrino singlet will be introduced in Section \ref{addsterile}. In this
section we will have large active -- sterile mixing but only the introduction
of a horizontal symmetry in Section \ref{discrete} will naturally restrict the
large mixing to the muon neutrino. In Section \ref{concl} we will present the
conclusions. 
\section{The Gelmini -- Roncadelli model with explicit lepton number
violation}
\label{Gelmini-Roncadelli}
In the GR model the Yukawa Lagrangian in the lepton sector is given by
\cite{GR} 
\begin{eqnarray}
\mathcal{L}_Y &=& \sum_{a,b} \left\{
\vphantom{\frac{1}{2}}
-c_{ab}\, \overline{\ell}_{aR} \phi^\dagger L_b 
\right. \nonumber\\
&+& 
\left.
\frac{1}{2} f_{ab} L_a^T C^{-1} i\tau_2 \Delta L_b 
\right\} + \mathrm{h.c.} \,, \label{yukawa}
\end{eqnarray}
where $a,b = 1,2,3$ are the summation indices over the active neutrino
degrees of freedom, $L_a$, $\ell_{aR}$ and $\phi$ denote the
left-handed lepton doublets, the right-handed lepton singlets and the
Higgs doublet, respectively. The Higgs triplet $\Delta$ is
represented in the form of a 2$\times$2 matrix. The coupling matrix for
the Higgs triplet is symmetric, i.e., $f_{ab} = f_{ba}$. Under 
$U \in \mbox{SU(2)}$ these multiplets transform as 
\begin{equation}\label{su2}
L_a \to U L_a \,, \;
\ell_{aR} \to \ell_{aR} \,, \;
\phi \to U \phi \,, \;
\Delta \to U \Delta U^\dagger \,.
\end{equation}
Their U(1) transformation properties are determined by the hypercharges:
\begin{equation}
\begin{array}{l|cccc}
& L_a & \ell_{aR} & \phi & \Delta \\ \hline
Y & -1 & -2 & 1 & 2
\end{array}
\end{equation}
Note that we are using the indices $a,b$ instead of $\alpha, \beta$
(\ref{mixing}). The two sets of indices are identical in a basis where
the mass matrix of the charged leptons is diagonal. However, for
reasons to become clear later, we want to use the more general
notation. The VEVs of the Higgs multiplets consistent with electric
charge conservation are given by
\begin{equation}\label{vev}
\langle \phi \rangle_0 =
\frac{1}{\sqrt{2}} \left(\begin{array}{c} 0 \\ v \end{array}\right)
\quad \mbox{and} \quad
\langle \Delta \rangle_0 =
\left(\begin{array}{cc} 0 & 0 \\ v_T & 0 \end{array}\right) \,.
\end{equation}
The relation between the triplet $\vec{\Phi}$, the 2$\times$2 matrix
$\Delta$ and the charged scalars contained in the triplet is found as 
\begin{equation}\label{triplet}
\Delta = \vec{\Phi} \cdot \vec{\tau} =
\left(\begin{array}{cc}
H^+ & \sqrt{2}H^{++} \\ \sqrt{2}H^0 & -H^+ 
\end{array}\right)
\end{equation}
with 
\begin{equation}
\vec{\Phi} = \left(\begin{array}{c}
\frac{1}{\sqrt{2}}(H^0 + H^{++}) \\ \frac{-i}{\sqrt{2}}(H^0 - H^{++}) \\
H^+ \end{array} \right) \,.
\end{equation}
The matrices $\tau_j$ ($j=1,2,3$) are the Pauli matrices.
In Eq.~(\ref{vev}) we have set $\langle H^0 \rangle_0 = v_T/\sqrt{2}$.
The most general Higgs potential involving $\phi$ and $\Delta$ is
written as 
\begin{eqnarray}
\lefteqn{V(\phi,\Delta) =} \nonumber \\
&& \hphantom{+} a\, \phi^\dagger\phi + \frac{b}{2} \tr(\Delta^\dagger\Delta)
+c\, (\phi^\dagger\phi)^2 + \frac{d}{4} 
\left( \tr(\Delta^\dagger\Delta) \right)^2 \nonumber\\
&& + \frac{e-h}{2} \phi^\dagger\phi \tr(\Delta^\dagger\Delta)
+ \frac{f}{4} \tr(\Delta^\dagger\Delta^\dagger)\tr(\Delta\Delta) \nonumber\\
&& + h\, \phi^\dagger \Delta^\dagger \Delta \phi  
+ \left( t\, \phi^\dagger \Delta \tilde{\phi} +
\mbox{h.c.} \right) \,, \label{pot}
\end{eqnarray}
where $\tilde{\phi} \equiv i\tau_2 \phi^\ast$. If the lepton number is
assumed to be
conserved one has to assign lepton number $-2$ to the Higgs triplet and
0 to the Higgs doublet \cite{GR} (see Eq.~(\ref{yukawa})).
This lepton number is explicitly broken by the last term in the Higgs
potential (\ref{pot}). Otherwise, this Higgs potential agrees with the
one given in Ref.~\cite{GR} with the same definition of the coupling
constants. All parameters in the Higgs potential
are real except $t$ which is complex in general.

By performing a global U(1) transformation, $v$ can always be chosen
real and positive. Because of the $t$-term in the potential we do not have
a second global symmetry, the lepton number \cite{GR}, 
to make $v_T$ real. Furthermore, $t$ can also be complex and,
therefore, in general 
we write $t = |t|e^{i\omega}$ and
$v_T = w e^{i\gamma}$ with $w \equiv |v_T|$.
We assume that the following orders of
magnitude for the parameters in the potential hold:
\begin{equation}\label{oomagn}
a,\:b \sim v^2 \:;\quad
c,\: d,\: e,\: f,\: h \sim 1 \:; \quad |t|\ll v \,.
\end{equation}
The potential as a function of the VEVs is given by 
\begin{eqnarray}
V( \langle \phi \rangle_0 , \langle \Delta \rangle_0 ) & = & 
\frac{1}{2} a v^2 + \frac{1}{2} b w^2 + \frac{1}{4}c v^4 +
\frac{1}{4}d w^4 \nonumber \\
&+& \frac{1}{4} (e-h) v^2 w^2 + 
v^2w|t|\cos(\omega+\gamma) \,. \nonumber \\
&& \label{pot-vev}
\end{eqnarray}
It has to be minimized as a function of the three parameters
$v, w, \gamma$ in order to obtain the relations between the VEVs and the
parameters of the Higgs potential. Minimization with respect to
$\gamma$, the phase of $v_T$, involves only the last term in
Eq.~(\ref{pot-vev}) with the minimum at
$\omega + \gamma = \pi$ or
\begin{equation}\label{tvtphases}
v_T = -w e^{-i\omega} \quad \mbox{and} \quad v_T t = - w |t| \,.
\end{equation}
With this relation the other two minimum conditions are
\begin{eqnarray}
a + cv^2 + \frac{e-h}{2} w^2 - 2|t|w &=& 0 \,, 
\label{min1}\\
b + dw^2 + \frac{e-h}{2} v^2 - \frac{|t|}{w} v^2 &=& 0 \,.
\label{min2}
\end{eqnarray}
With the assumptions (\ref{oomagn}) we find the approximate solution
\begin{equation}
v^2 \simeq -\frac{a}{c} \quad \mbox{and} \quad 
w \simeq |t|\, \frac{v^2}{b + (e-h)v^2/2} \,.
\end{equation}
Thus we see that $w \sim |t|$, i.e., the triplet VEV is of the order
of the parameter $|t|$ in the Higgs potential. The fine-tuning 
to get a small triplet VEV is therefore simply given
by $|t| \ll v$, which should find an explanation in a more complete
theory which has the GR model as a low energy
limit.\footnote{Alternatively, one could use $b \gg v^2$ to get a
small triplet VEV \cite{ma}.} This is the analoguous
situation as with the Standard Model and the see-saw mechanism for 
light neutrino masses, where the large mass scale of the right-handed
neutrino singlets is assumed to come, e.g., from Grand Unification.

Eqs.~(\ref{yukawa}) and (\ref{vev}) give rise to the mass terms for the charged
leptons and the neutrinos:
\begin{eqnarray}
- \left( \overline{\ell}_R \mathcal{M}_\ell \ell_L
+ \mathrm{h.c.} \right)
&& \; \mbox{with} \quad
\mathcal{M}_\ell = \frac{v}{\sqrt{2}} \left(c_{ab}\right) \,,
\label{chleptmass}\\
\frac{1}{2} \nu^T_L C^{-1} \mathcal{M}_\nu \nu_L
+ \mathrm{h.c.} \hphantom{)}
&& \; \mbox{with} \quad
\mathcal{M}_\nu = v_T \left( f_{ab} \right) \,.
\label{numass}
\end{eqnarray}

As mentioned earlier, if the cubic
term in the potential (\ref{pot}) is absent, then
there are two independent symmetries,
the gauge group and the lepton number, which allow us to adopt the
convention $v$ and $v_T$ both real and positive. This means that in
the Higgs sector CP cannot be broken. It could, of course, be violated
explicitly by complex Yukawa couplings. In the presence of
the cubic term the situation is more complicated. We define a CP
transformation 
\begin{equation}\label{CP2}
\phi \to \phi^\ast \,, \quad
\Delta \to \rho \Delta^\ast \quad \mbox{with} \quad |\rho| = 1
\end{equation}
for the two scalar multiplets. Invariance of the Higgs potential under
this CP transformation leads to
the condition
\begin{equation}\label{CPV}
t^\ast = \rho t
\end{equation}
for the parameter $t$. Interpreted in another way, for any complex
phase $\omega$ of $t$, the Higgs potential is invariant under the CP
transformation (\ref{CP2}) if we choose
\begin{equation}\label{rho}
\rho = e^{-2i\omega} \,.
\end{equation}
Let us check that the VEVs are indeed invariant under
the CP symmetry defined by Eqs.~(\ref{CP2}) and (\ref{rho}). This is
clear for $\langle \phi \rangle_0$ since $v$ is real. Taking into
account that the phase of $v_T$ is given by Eq.~(\ref{tvtphases}) at
the minimum of the potential and using Eqs.~(\ref{CP2}) and
(\ref{rho}) we find
\begin{equation}
\langle \Delta \rangle_0 =
\left(\begin{array}{cc} 0&0 \\ v_T & 0 \end{array}\right) 
\stackrel{\mathrm{CP}}{\longrightarrow} 
\rho \langle \Delta \rangle_0^* = 
\rho \left( \begin{array}{cc} 0 & 0 \\ v_T^\ast &0 \end{array}\right) =
\left( \begin{array}{cc} 0&0 \\ v_T &0 \end{array}\right) \,.
\end{equation}
Hence we see that the vacuum state is invariant under CP, 
regardless of the complex phase of $t$
in the Higgs potential (\ref{pot}) \cite{CP}, 
and thus CP cannot be spontaneously broken.
Extending the CP transformation (\ref{CP2}) by
\begin{equation}\label{CP1}
\Psi(x^0,\vec{x}) \to -C \Psi^\ast(x^0,-\vec{x})
\end{equation}
for the fermionic multiplets and assuming that
the vector bosons transform in the usual way,
we obtain the conditions
\begin{equation}\label{CPcond}
c_{ab} = c_{ab}^\ast \,, \quad
-\rho f_{ab} = f_{ab}^\ast 
\end{equation}
for CP invariance of the fermionic Lagrangian.
Using the second relation in Eq.~(\ref{CPcond}) we find with
Eq.~(\ref{rho}) that
\begin{equation}\label{CPf}
f_{ab}^\ast = -e^{-2i\omega} f_{ab} \,.
\end{equation}
If we define $f'_{ab}$ by
\begin{equation}\label{deff'}
f'_{ab} = ie^{-i\omega} f_{ab}
\end{equation}
then Eq.~(\ref{CPf}) implies 
\begin{equation}\label{wf'}
f'_{ab} \in \mathbf{R} \quad \mbox{and} \quad
v_T f_{ab} = i w f'_{ab} \,.
\end{equation}
In the following we will assume CP invariance for simplicity, though
it is not essential for the construction of our model.

In the GR model the relation between the $W$ and $Z^0$ masses is
obtained as \cite{GR,erler}
\begin{equation}
\frac{M_W^2}{M_Z^2 \cos^2\theta_W} = 
\frac{1 + 2w^2/v^2}{1 + 4w^2/v^2} \,,
\end{equation}
whereas in the Standard Model this ratio is 1. From precision data, in
Ref.~\cite{erler} the bound
\begin{equation}\label{bound}
\frac{w}{v} \lesssim 0.03
\end{equation}
was obtained at 95\% CL. If there are several Higgs doublets with VEVs
$v_k$ then $v$ has to replaced by $(\sum_k |v_k|^2)^{1/2}$ in
Eq.~(\ref{bound}). 

With the definitions
\begin{equation}\label{decomp}
\phi^0 = \frac{1}{\sqrt{2}} ( v + \varphi_R + i \varphi_I ) 
\,, \quad
H^0 = \frac{1}{\sqrt{2}} e^{i\gamma}(w + H_R + i H_I) \,,
\end{equation}
where the scalar fields with the subscripts $R$ and $I$ are real fields, we 
write the couplings of the neutral scalars to the $Z^0$ boson as 
\begin{eqnarray}
\frac{\sqrt{g^2+g'^2}}{2} & Z^\mu &
\left\{ (\partial_\mu \varphi_R)\varphi_I - 
(\partial_\mu \varphi_I)(\varphi_R + v) \right. \nonumber \\
& + & \left.
2(\partial_\mu H_R)H_I - 2(\partial_\mu H_I)(H_R + w) \right\} \,.
\nonumber \\ && \label{Z} 
\end{eqnarray}
The quantities $g$ and $g'$ are the gauge coupling constants of
SU(2)$\times$U(1). Note that the linear combination
\begin{equation}\label{pseudoZ}
2w H_I + v \phi_I 
\end{equation}
in Eq.~(\ref{Z}) is proportional to the pseudo-Goldstone boson field
associated with the $Z^0$.

CP invariance of the scalar sector under the transformation
(\ref{CP2}) and the decomposition (\ref{decomp}) show that the
$R$-fields are CP-even and the $I$-fields CP-odd. Therefore,
the mass matrix of the four real scalar fields splits into
two separate 2$\times$2 matrices for the real and imaginary parts, i.e., 
\begin{equation}
\mathcal{L}^0_S = -
\frac{1}{2} (H_R, \phi_R) \mathcal{M}^2_R 
\left( \begin{array}{c} H_R \\ \phi_R \end{array} \right) -
\frac{1}{2} (H_I, \phi_I) \mathcal{M}^2_I 
\left( \begin{array}{c} H_I \\ \phi_I \end{array} \right) \,.
\end{equation}
Using the minimum conditions
Eqs.~(\ref{min1}) and (\ref{min2}) we get
\begin{eqnarray}
\mathcal{M}^2_R &=& \left(\begin{array}{cc}
2dw^2 + q v^2 & (e - h - 2q) w v  \\
(e - h - 2q)w v  &  2 cv^2
\end{array}\right) \,, \label{mR} \\
\mathcal{M}^2_I &=& q \left(\begin{array}{cc}
v^2 & -2w v \\ -2w v & 4w^2 \end{array}\right) \,, \label{mI}
\end{eqnarray}
where we have defined
\begin{equation}\label{defq} 
q \equiv |t|/w \,,
\end{equation}
which is a positive quantity of order one according to the assumptions
(\ref{oomagn}). 
The eigenvalues of the matrices 
$\mathcal{M}^2_R$ and $\mathcal{M}^2_I$ are given by
\begin{eqnarray}
\label{mR1}
m^2_{R1} & \simeq & 
2cv^2 + \frac{(e - h - 2q)^2}{2c - q}w^2 \,, \\
\label{mR2}
m^2_{R2} & \simeq &
qv^2 - \left[ \frac{(e - h - 2q)^2}{2c - q} - 2d \right]w^2 \,, \\
\label{mI1}
m^2_{I1} & = & q(v^2 + 4w^2) \,, \\ 
\label{mI2}
m^2_{I2} & = & 0 \,,
\end{eqnarray}
respectively.
The masses of the $R$-fields are given up to first order in
$w^2$, whereas the masses of the $I$-fields are exact.
The zero eigenvalue corresponds to the linear combination Eq.~(\ref{pseudoZ}).
In the GR model without the cubic term in the Higgs potential,
we have $t=0$ (or $q=0$) and also the second eigenvalue of the $I$-fields
is zero. This eigenvalue corresponds to the Goldstone boson (majoron)
which results from the spontaneous
breaking of the U(1) symmetry connected with lepton number conservation.
Moreover, in this case $m^2_{R2}$ is of order $w^2$ and, therefore,
the $Z^0$ can decay into the majoron and the light scalar with a decay width 
of two neutrino flavours \cite{georgi}. Thus, 
the GR model is ruled out because of the LEP results. 
Eqs.~(\ref{mR2}) and (\ref{mI1}) show that with $q$ of order one
all physical neutral scalars can be made heavy enough such that
the $Z^0$ cannot decay into them. Consequently, the GR model with a
cubic term in the Higgs potential is consistent
with the LEP data \cite{ma}.

For completeness we also mention the masses of the charged scalars.
The mass Lagrangian for the singly char\-ged scalars is given by
\begin{equation}
\mathcal{L}^\pm_S = -
(H^-, \phi^-) \mathcal{M}_+^2 
\left( \begin{array}{c} H^+ \\ \phi^+ \end{array} \right)
\end{equation}
with
\begin{equation}
\mathcal{M}_+^2 = \left(\begin{array}{cc}
2(q + h/2)w^2   & \sqrt{2} v (t^\ast - v_T h/2) \\
\sqrt{2} v (t - v_T^\ast h/2) & (q+h/2)v^2 
\end{array}\right) \,.
\end{equation}
The field $\phi^+$ denotes the charged component of the scalar doublet.
One mass eigenvalue of this matrix is zero corresponding to the
pseudo-Goldstone boson which gives mass to the $W$ boson. The mass of
the single physical scalar with charge +1 is computed as 
\begin{equation}
m^2_+ = \left(q + \frac{h}{2} \right) (v^2 + 2w^2) \,.
\end{equation}
For the mass of the scalar with charge +2 one finds
\begin{equation}
m^2_{H^{++}} = (h + q) v^2 + 2fw^2 \,.
\end{equation}
Note that, as expected, all physical charged scalars are heavy 
regardless if we set
$t=0$ or not and hence the $Z^0$ cannot decay into charged Higgses for
$h \sim 1$.
\section{Adding a sterile neutrino}
\label{addsterile}
Adding a fourth neutrino to the GR model with a cubic term in the Higgs
potential, we have to take into account that
because of the LEP measurements of the $Z^0$ decay width this neutrino
must not couple to the $Z^0$. So it has to be a
trivial singlet under SU(2)$\times$U(1). Since it has no gauge interactions
it is called a sterile neutrino. In analogy with the fields
$\ell_{aR}$ we denote it by the right-handed field $\nu_{sR}$.
The only new gauge invariant terms involving the sterile neutrino
field are given by
\begin{equation}\label{newterms}
\left( - \sum_a h_a \overline{\nu}_{sR}
\tilde{\phi}^\dagger L_a + \frac{1}{2}M_s \, \nu_{sR}^T C^{-1} \nu_{sR}
\right) + \mbox{h.c.}
\end{equation}
The Majorana mass $|M_s|$ of the sterile neutrino is usually assumed
to be much larger than the other neutrino masses and 
could typically be of the order of the GUT scale. Therefore, we opt
for introducing a symmetry forbidding the mass term in
Eq.~(\ref{newterms}). It turns out, however, that it is not possible 
to construct such a symmetry, by assigning phase factors to all the
multiplets of the model, without forbidding other crucial terms of the
model like the cubic term in the Higgs potential. This forces us to
introduce a second scalar doublet $\phi_s$. Then we can conveniently 
define a symmetry $S$ by
\begin{equation}\label{S}
S: \quad \nu_{sR} \to e^{i\alpha} \nu_{sR},\quad
\phi_s \to e^{i\alpha} \phi_s \,.
\end{equation}
All other multiplets transform trivially. $S$ forbids the
Majorana mass term in Eq.~(\ref{newterms}) provided 
$e^{2i\alpha} \neq 1$. Now instead of
Eq.~(\ref{newterms}) we have
\begin{equation}\label{new}
- \left( \sum_a h_a \overline{\nu}_{sR}
\tilde{\phi}_s^\dagger L_a  + \mbox{h.c.} \right) \,.
\end{equation}

After spontaneous symmetry breaking Eq.~(\ref{new}) gives the mass term
\begin{equation}\label{newmass}
- \left( \frac{v_s}{\sqrt{2}} \sum_a h_a \overline{\nu}_{sR} \nu_{aL} 
+ \mbox{h.c.} \right) \; \mbox{where} \;\,
\langle \phi_s \rangle_0 = \frac{1}{\sqrt{2}}
\left(\begin{array}{c} 0 \\ v_s \end{array}\right) \,.
\end{equation}
It is easily seen that the condition $e^{2i\alpha} \neq 1$ causes the
Higgs potential to be invariant under $S$ (\ref{S}) interpreted as a
continuous symmetry with $\alpha \in \mathbf{R}$. Therefore,
by spontaneous symmetry breaking we obtain a Goldstone boson. One can
show with the methods of Section \ref{Gelmini-Roncadelli} that the
other scalars are heavy and thus there is no contradiction with the
measurement of the $Z^0$ width. The couplings of the Goldstone boson are
similar to those in the model of Ref.~\cite{chikashige} (see also
\cite{GR}) which was shown to be compatible with experimental data.
The continuous symmetry $S$ allows to choose $v_s > 0$ and $\phi_s$
transforms like $\phi$ (\ref{CP2}) under CP.
Then the invariance of the term (\ref{new}) under CP implies
$h_a^\ast = h_a$.

The mass terms Eqs.~(\ref{numass}) and (\ref{new}) are combined in a
4-neutrino Majorana mass term as
\begin{equation}\label{4nuL}
\frac{1}{2} \left( \nu_L^T , \nu_{sL}^T \right) 
C^{-1} \mathcal{M}_{4\nu}
\left(\begin{array}{c} \nu_L \\ \nu_{sL} \end{array}\right)
+ \mbox{h.c.}
\end{equation}
with 
\begin{equation}\label{4nu}
\mathcal{M}_{4\nu} =
\left(\begin{array}{cc} i w F & \frac{v_s}{\sqrt{2}}h^T \\
\frac{v_s}{\sqrt{2}}h & 0 \end{array}\right) \,,
\end{equation}
where we have defined the charge-conjugate field 
$\nu_{sL} \equiv (\nu_{sR})^c$, 
the 3$\times$3 matrix $F \equiv (f'_{ab})$ and the line vector 
$h \equiv (h_a)$. Now we fix the notation of the diagonalizing matrices of the
mass terms. The mass matrix of the charged leptons (see
Eq.~(\ref{chleptmass})) and of the neutrinos (see Eq.~(\ref{4nu}))
are diagonalized by
\begin{equation}\label{diag}
W_\ell^\dagger \mathcal{M}_\ell V_\ell = \hat{\mathcal{M}}_\ell
\quad \mbox{and} \quad
V_\nu^T \mathcal{M}_{4\nu} V_\nu = \hat{\mathcal{M}}_{4\nu} \,,
\end{equation}
respectively. From $V_\ell$ and $V_\nu$ the mixing matrix (\ref{mixing}) is
computed as
\begin{equation}\label{U}
U = {V'_\ell}^\dagger V_\nu \quad \mbox{with} \quad
V'_\ell = \left(\begin{array}{cc} V_\ell & 0 \\ 0 & 1 \end{array} \right) \,.
\end{equation}

For the further discussion we will stick to the following order
of magnitude assumptions:
\begin{equation}\label{oomagn1}
F \sim h \quad \mbox{and} \quad v \sim v_s \sim 100 \: \mbox{GeV} \,.
\end{equation}
This makes our mass matrix (\ref{4nu}) analoguous to the one 
obtained in Ref.~\cite{CKL} with the singular see-saw mechanism. With
Eqs.~(\ref{oomagn1}) and (\ref{vev-rel}), the elements in the 
mass matrix (\ref{4nu}) are of two different orders of magnitude, represented 
by the VEVs $v_s$ and $|v_T|$ or 
$\mu \sim w |f'_{ab}| \ll M \sim v_s |h_a|$ $\forall a,b$.
With the ordering $m_1 < m_2 < m_3 < m_4$ of the
neutrino masses, repeating the arguments of Ref.~\cite{CKL}, we read off from
Eq.~(\ref{4nu}) that 
\begin{equation}\label{4numasses}
m_1, \: m_2 \sim \mu, \quad m_3, \: m_4 \sim M, \quad m_4-m_3 \sim \mu,
\end{equation}
and with the definition $\Delta m^2_{jk} = m^2_j-m^2_k$ we obtain
\begin{equation}
\Delta m^2_{21} \sim \mu^2, \quad \Delta m^2_{43} \sim \mu M, \quad
\Delta m^2_{41} \sim M^2 \,.
\end{equation}
Therefore,
in a natural way three different scales for the mass-squared differences
occur. If we set $\Delta m^2_{21} = \Delta m^2_{\mathrm{solar}} \sim
10^{-5}$ eV$^2$ and $\Delta m^2_{41} = \Delta m^2_{\mathrm{LSND}} \sim
1$ eV$^2$ we get $\Delta m^2_{43} \sim 3\times 10^{-3}$ eV$^2$,
which is just the right order of magnitude for $\Delta
m^2_{\mathrm{atm}}$. In this way we obtained the mass spectrum of
Scheme B (\ref{AB}), which forces us to envisage 
$\nu_e\to\nu_\tau$ MSW transitions as a
solution for the solar neutrino deficit and $\nu_\mu\to\nu_s$
transitions to explain the atmospheric neutrino anomaly.
The ratio $\mu / M \sim |v_T| / v_s \sim 3 \times 10^{-3}$ 
is well below the constraint (\ref{bound}). Note that a solution
of the solar neutrino problem by vacuum oscillations with 
$\Delta m^2_{\mathrm{solar}} \sim 10^{-10}$ eV$^2$ is not possible in
the scenario discussed here.

Finally we want to remark that with the assumptions (\ref{oomagn1}) 
the elements of $F$ and $h$ must be very small: 
if we want $M \sim v_s h$ to be
of order 1 eV, then $v_s \sim 100$ GeV implies that $F$ and 
$h$ must be of order $10^{-11}$. However, with
all coefficents in $F$ and $h$ being of the same order of magnitude,
the structure of the 4-neutrino mass spectrum corresponding to Scheme B
is obtained in a natural way, simply by having the two scales given by
$v_s$ and $|v_T|$.
\section{A discrete symmetry to implement large $\nu_\mu$--$\nu_s$
mixing}
\label{discrete}
The shortcomings of the model discussed in the previous section and in
Ref.~\cite{CKL}, which were also noticed in Ref.~\cite{liu}, are that one 
still has to resort to fine-tuning in order to specify the large active --
sterile 
neutrino mixing to large $\nu_\mu$--$\nu_s$ mixing and also to get the correct
small $\nu_e$--$\nu_\mu$ mixing as required by the result of the LSND
experiment \cite{LSND}. In the following we propose a symmetry called $T$ 
which replaces the symmetry $S$ of the previous section and 
removes the first shortcoming. 
It requires us, however, to enlarge the Higgs content
of the scenario in the previous section by an additional scalar doublet. This
will allow us to give also a plausible reason for the small $\nu_e$--$\nu_\mu$
mixing. 

In order to implement large $\nu_\mu$--$\nu_s$ mixing we require that in the
Lagrangian (\ref{new}) the right-handed neutrino singlet couples to only one
left-handed lepton doublet which we denote by $L_3$. As we shall see, the
non-trivial transformation of the left-handed lepton doublets under $T$
necessitates the introduction of two scalar doublets $\phi_{1,2}$ 
in the Lagrangian (\ref{yukawa}) in order to have only
non-zero charged lepton masses.
The symmetry $T$ is defined via the prescription
\begin{eqnarray}
T: & \quad \nu_{sR} \to i \nu_{sR} \:, &
\quad \phi_s \to -i \phi_s \:, \nonumber \\
& \quad \phi_2 \to - \phi_2 \:, &
\quad L_3 \to - L_3 \:. \label{T}
\end{eqnarray}
All other fields transform trivially under $T$. 
Taking into account $T$, the Yukawa
couplings for the two Higgs doublets $\phi_{1,2}$ are given by
\begin{equation}\label{yukawa1}
- \left\{ \left( \sum_{a=1}^3 \sum_{b=1}^2
c_{ab} \overline{\ell}_{aR} \phi_1^\dagger L_b
+ \sum_{a=1,2,3} y_a
\overline{\ell}_{aR} \phi_2^\dagger L_3 \right)
+ \mbox{h.c.} \right\} \,.
\end{equation}
With the three Higgs doublets $\phi_{1,2,s}$ we have the terms
\begin{equation}\label{term}
\phi_s^\dagger \phi_1 \phi_s^\dagger \phi_2 
\quad \mbox{and} \quad
\phi_1^\dagger \phi_2 \phi_1^\dagger \phi_2 
\end{equation}
in the Higgs potential. As a consequence, the only U(1) allowed by the
potential is the one associated with the hypercharge.
Thus with the symmetry $T$ we forbid a
Majorana mass term of the right-handed neutrino singlet and avoid also a
Goldstone boson at the same time.

Defining $\langle \phi_k^0 \rangle = v_k/\sqrt{2}$ ($k=1,2,s$), we assume
that all doublet VEVs are of the same order of magnitude. Now with 
the two cubic terms pertaining to $\phi_{1,2}$ and the quartic 
terms (\ref{term}) in the Higgs potential, CP can be broken explicitly or
spontaneously in the Higgs sector. In the following we will stick to
CP conservation and assume real VEVs for simplicity.
The Yukawa couplings (\ref{yukawa1}) give the mass matrix for the
char\-ged leptons
\begin{equation}\label{leptmassmatr}
\mathcal{M}_\ell = \left( \frac{v_1}{\sqrt{2}} 
\left( \begin{array}{cc}
c_{11} & c_{12} \\ c_{21} & c_{22} \\ c_{31} & c_{32}
\end{array}\right) \!, \frac{v_2}{\sqrt{2}}
\left(\begin{array}{c} y_1 \\ y_2 \\ y_3 \end{array}\right) \right).
\end{equation}
From this equation it is obvious that a third scalar doublet $\phi_2$
is needed to reproduce the charged lepton mass spectrum.
Because of the symmetry $T$ the neutrino mass matrix splits 
into two $2\times 2$ matrices:
\begin{equation}\label{numassmatr}
\mathcal{M}_{4\nu} = \left(\begin{array}{cc}
\mathcal{M}_{12} & 0 \\ 0 & \mathcal{M}_{3s} \end{array}\right)
\end{equation}
with
\begin{equation}\label{numass2}
\mathcal{M}_{12} = iw \left(\begin{array}{cc}
f'_{11} & f'_{12} \\ f'_{12} & f'_{22}
\end{array}\right) \quad \mbox{and} \quad
\mathcal{M}_{3s} = \left(\begin{array}{cc}
iwf'_{33} & \frac{v_s}{\sqrt{2}} h_3 \\
\frac{v_s}{\sqrt{2}} h_3 & 0
\end{array}\right).
\end{equation}
Let us consider the matrix $\mathcal{M}_{3s}$. Up to order $w$ it
gives the neutrino masses
\begin{equation}\label{mass3s}
\frac{1}{\sqrt{2}} |v_s h_3| \pm \frac{1}{2} w f'_{33}
\end{equation}
and a mixing angle $\theta_{3s}$ obtained by
\begin{equation}\label{mixing3s}
\sin^2 2\theta_{3s} \simeq 1 - \frac{1}{2}
\left( \frac{w f'_{33}}{v_s h_3} \right)^2 \,.
\end{equation}

With $v_s \sim v_{1,2}$, $f'_{ab} \sim h_3$ (\ref{oomagn1}) 
and Eq.~(\ref{bound}), $\sin^2 2\theta_{3s}$ is
1 for all practical purposes and, naturally, we want to
associate the matrix $\mathcal{M}_{3s}$ with the $\nu_\mu - \nu_s$ solution
of the atmospheric neutrino problem. Furthermore, the other 
$2 \times 2$ mass matrix $\mathcal{M}_{12}$ has all matrix elements of
the same order of magnitude and, therefore, suggests to explain
the solar neutrino problem by $\nu_e - \nu_\tau$
oscillations with the large angle MSW solution. 

The diagonalization matrix $V_\nu$ of the neutrino mass matrix
(\ref{numassmatr}) consists of two $2 \times 2$ submatrices, i.e.,
\begin{equation}\label{V}
V_\nu = \left( \begin{array}{cc} V_{12} & 0 \\
0 & V_{3s}     \end{array} \right) \,.
\end{equation}
So $V_\nu$ does not have, e.g., $\nu_e$--$\nu_\mu$ mixing necessary
to describe the LSND experiment, provided we associate the submatrices
of $\mathcal{M}_\nu$ (\ref{numassmatr}) with neutrino flavours as done
in the previous paragraph. However, in order to
obtain the mixing matrix $U$ we have to multiply
$V_\nu$ with ${V'_\ell}^\dagger$ (see Eqs.~(\ref{mixing}) and 
(\ref{U})) which is
determined by the diagonalization of $\mathcal{M}_\ell$
(\ref{chleptmass}). Our model does not specify $\mathcal{M}_\ell$. In
order to proceed further we make the following assumption regarding
$V_\ell$: In analogy with the quark sector we assume that $V_\ell$ is
close to a diagonal phase matrix. This amounts to 
$|(V_\ell)_{1e}| \simeq |(V_\ell)_{2\tau}| \simeq |(V_\ell)_{3\mu}|
\simeq 1$, since these elements correspond to the diagonal elements of
$V_\ell$ in our model. All other elements are assumed to be small.

Clearly, this assumption is in agreement with the scenarios for the
atmospheric and solar neutrinos proposed above. Let us now discuss how
the result of the LSND experiment fits into the model. This experiment
measures the short-baseline transition amplitude
\begin{equation}
P^{(\mathrm{SBL})}_{\bar\nu_\mu\to\bar\nu_e} =
A_{e;\mu} \sin^2 \frac{\Delta m^2_{41}L}{4E_\nu} \,,
\end{equation}
where $L$ is the distance between the neutrino source and detector,
$E_\nu$ is the neutrino energy and the oscillation amplitude
$A_{e;\mu}$ is obtained from the mixing matrix as
\begin{equation}\label{A}
A_{e;\mu} = 4\, \left| \sum_{j=3,4} U_{ej}^\ast U_{\mu j} \right|^2 \,.
\end{equation}
Considering the structure (\ref{V}) of $V_\nu$ one finds
\begin{equation}\label{Aexp}
A_{e;\mu} = 4\, |(V_{\ell})_{3e}|^2 |(V_{\ell})_{3\mu}|^2 \simeq
4\, |(V_{\ell})_{3e}|^2 \,.
\end{equation}
In the last step we have used our assumption about $V_\ell$. 
The experimental result of the LSND experiment, taking into account
other short-baseline experiments which have seen no indication in
favour of neutrino oscillations, is expressed as \cite{LSND}
\begin{equation}\label{LSND}
2 \times 10^{-3} \lesssim A_{e;\mu} \lesssim 3 \times 10^{-2} \,,
\end{equation}
where the bounds result from the LSND-allowed region (90\% CL).
Thus, from Eqs.~(\ref{A}) and (\ref{Aexp}) it follows that
$|(V_{\ell})_{3e}|$ is of the order of $10^{-2}$ to $10^{-1}$
conforming with the above assumption as expected.

To conclude this section we want to make some remarks about the
scalars. Now there are two cubic terms corresponding to $\phi_{1,2}$
and, therefore, 
two coupling constants $t_{1,2}$ in the potential (see
Eq.~(\ref{pot})) which must both be much smaller than the doublet VEVs
and of the order of the triplet VEV. The assumption of CP conservation
simplifies the discussion of the neutral scalar masses because it
causes the 8$\times$8 scalar mass matrix to split into two 
4$\times$4 mass matrices, one for the $R$-fields and one for the
$I$-fields (see Section \ref{intro}). One can again show that all
physcial neutral scalars are heavy of the order of the doublet
VEVs. The same is true for the charged scalars. 
\section{Conclusions}
\label{concl}
In this paper we have constructed a 4-neutrino model based on the Gelmini --
Roncadelli model, which extends the Standard Model by a scalar triplet
$\Delta$ leading to Majorana neutrino masses at the tree level. 
In order to prevent the $Z^0$ decay into light neutral scalars we have
explicitly broken the lepton number of the original GR model by a
cubic term in the Higgs potential. We have introduced a sterile
neutrino and coupled it to the standard lepton gauge doublets by a
separate Higgs doublet $\phi_s$. It is well known that the triplet VEV
must be much smaller than the doublet VEVs because of the tree level
relation $M_W = M_Z \cos \theta_W$. One of the main points of our model is to
exploit the presence of the two scales represented by the triplet and
doublet VEVs. In this way, assuming that the $\Delta$ and $\phi_s$
couplings are of the same order of magnitude, we immediately arrive at
a model which reproduces the neutrino mass spectrum of Scheme B
(\ref{AB}), 
one of the two schemes allowed by all present neutrino oscillation
data. This model, described in Section \ref{addsterile}, is completely
analoguous to the model of Ref.~\cite{CKL} which invokes the singular
see-saw mechanism. However, in this case the heavy scale of the
see-saw mechanism is quite low of the order of keV. Our model avoids
this -- we have only four light neutrinos -- at the expense of the
triplet VEV being much smaller than the the doublet VEVs occurring in the
model. Of course, the smallness of the triplet VEV can only be
obtained by fine-tuning in the Higgs potential and the hope is that in
a more complete theory this problem of fine-tuning is resolved.

The scenario of Section \ref{addsterile} automatically leads to a
large active -- sterile neutrino mixing. However, any linear
combination of the active neutrinos could have this large mixing. In
Section \ref{discrete} we have introduced a symmetry which splits the
$4 \times 4$ Majorana neutrino mass matrix into two $2 \times 2$
matrices. The diagonalization matrices of both $2 \times 2$ matrices
contain a large angle, one of them is $\pi/4$ for all practical
purposes. In this version of the model we need three Higgs doublets.
Neglecting for a moment the part of the mixing matrix $U$ coming from
the charged lepton sector (see Eq.~(\ref{U})), the mixing matrix also
separates into two $2 \times 2$ matrices. In this way we naturally
obtain a model where the solar neutrino problem is explained by 
large mixing angle MSW $\nu_e\to\nu_\tau$ transitions and the
atmospheric neutrino problem by $\nu_\mu\to\nu_s$ transitions with
mixing angle $\pi/4$.
With the assumption that in the charged lepton sector the left-handed
diagonalization matrix of the mass matrix is close to a diagonal phase matrix,
the scenario just described is not very much disturbed. Moreover,
one can exploit $V_\ell$ (\ref{diag}) to incorporate the LSND result
of small $\nu_e$--$\nu_\mu$ mixing, which is forbidden if $V_\ell$
is diagonal. 

This assumption about the charged lepton sector is certainly a weak
point of our model, but, in any case we have no explanation for the
charged lepton spectrum either. Furthermore, the assumption of equal
order of magnitude of the $\Delta$ and $\phi_s$ couplings leads to
very small coupling constants of order $10^{-11}$ to obtain the
smallness of the neutrino masses relative to $M_W$ and $M_Z$.
Also this has to find a natural explanation in a larger theory. 
Despite of these shortcomings, we want to stress that our model
only requires the minimal extension of the fermionic sector of the Standard
Model necessary for a 4-neutrino scheme and that
looking for an explanation of the 4-neutrino mass spectrum indicated by the
experimental data
in terms of VEVs of scalar multiplets could provide interesting clues
for theories with scales beyond the gauge boson masses of the Standard Model.

\end{document}